\documentclass[twocolumn,reprint,amsmath,amssymb,aps,prl,nofootinbib]{revtex4-2}

\usepackage{graphicx}
\usepackage{tikz}
\makeatletter\let\oldincludegraphics\includegraphics
\renewcommand{\includegraphics}[2][]{\IfFileExists{#2}{\oldincludegraphics[#1]{#2}}{\fbox{\parbox[c][2.6cm][c]{0.8\linewidth}{\centering\small figure file missing:\\\texttt{\detokenize{#2}}}}}}\makeatother
\usepackage{braket}
\usepackage{bm}
\usepackage{mathrsfs}
\usepackage{xcolor}
\usepackage{multirow}
\usepackage[hidelinks]{hyperref}
\usepackage{verbatim}
\usepackage[normalem]{ulem}

\newcommand{\Nseg}{N_{\rm seg}}

\begin{document}

\title{Constructing likelihoods for a non-Gaussian
gravitational-wave background}

\author{T\"ore Boybeyi}
\email{boybe001@umn.edu}
\affiliation{School of Physics and Astronomy, University of Minnesota, Minneapolis, MN 55455, USA}
\author{Vuk Mandic}
\email{vuk@umn.edu}
\affiliation{School of Physics and Astronomy, University of Minnesota, Minneapolis, MN 55455, USA}
\author{Joseph D. Romano}
\email{joseph.romano@utrgv.edu}
\affiliation{Department of Physics and Astronomy, University of Texas Rio Grande Valley, One West University Boulevard, Brownsville, TX 78520, USA}

\date{\today}

\begin{abstract}
The stochastic gravitational-wave background from stellar-mass compact binaries is non-Gaussian since it is made up of a finite number of individual mergers. 
Over the years, several methods have been developed to search for this background beyond the standard cross-correlation statistic. 
We reformulate the data analysis problem for such astrophysical backgrounds in terms of the characteristic function of the data in multiple detectors, and show that the existing searches arise as different limiting cases. 
We also show that \textcolor{black}{new} searches can be constructed through different approximations of the characteristic function, which gives a concrete way to engineer \textcolor{black}{suitable} approximate sufficient statistics. 
To illustrate this framework in the context of a simple example, we use the formalism proposed here to analyze an exactly solvable toy model.
\end{abstract}

\maketitle

{\it Introduction}---
\label{sec:intro}
A stochastic gravitational-wave background (GWB) is the incoherent superposition of gravitational waves (GWs) produced by a large number of individually unresolvable sources~\cite{Allen:1999,Romano:2017}.
These sources can be either astrophysical or cosmological in origin~\cite{Caprini:2018mtu,LIGOScientific:2025kry,Regimbau:2011rp}.
An astrophysical GWB  produced by compact binaries is a prime target of searches with the LIGO, Virgo, and KAGRA (LVK) detectors~\cite{LIGOScientific:2014pky,VIRGO:2014yos,KAGRA:2020tym,Virgo:2025aai,LIGOScientific:2026mjf}. 
It is also a target of future ground- and space-based detectors such as the Einstein Telescope, Cosmic Explorer, and LISA~\cite{Evans:2021gyd, ET:2025xjr, Punturo:2010zz,LISA:2017pwj,LISA:2024hlh}. 
In the nanohertz band, pulsar-timing arrays have recently reported marginal to strong evidence for a GWB~\cite{NANOGrav:2023gor,EPTA:2023fyk,Reardon:2023gzh,Xu:2023wog,IPTA:2023}, with the population of supermassive black-hole (SMBH) binaries in the centers of merging galaxies being a likely source~\cite{Agazie:2024jbf,NANOGrav:2023hfp}.

Here, we focus on ground-based detectors, whose relevant astrophysical sources are coalescing binary black holes (BBHs), binary neutron stars (BNSs) and neutron star black hole binaries (NSBHs). 
These populations differ qualitatively.
BBH mergers are more massive and shorter-lived in band, so their signals are {\it intermittent} (or ``popcorn''-like)~\cite{Meacher:2015iua,LIGOScientific:2016fpe}. 
BNS signals spend much longer in the sensitive band of the detectors and tend to overlap, forming a near-continuous confusion background~\cite{Regimbau:2009rk,Jenkins:2019uzp,Cornish:2015pda,LIGOScientific:2017zlf}.

Historically, the GWB has been approximated as a Gaussian random field~\cite{Allen:1999}. 
This approximation is valid for a background of cosmological origin, being the sum of contributions from many independent patches, thus obeying the central-limit theorem. 
For a Gaussian random field, the two-point correlation of the data is a sufficient statistic.
The GWB is then fully characterized by its second moment, and the target of a search collapses onto the dimensionless energy-density spectrum 
\begin{equation}
\Omega_{\rm GW}(f)
  \equiv \frac{1}{\rho_c}\frac{d\rho_{\rm gw}}{d\ln f}
   = \frac{2\pi^2}{3H_0^2}\,f^3\,S_h(f)\,,
\label{eq:omega}
\end{equation}
where $S_h(f)$ is its strain power spectral density and $\rho_c\equiv 3 c^2 H_0^2/8\pi G$ is the critical density needed to close the universe.

Standard LVK analyses cross-correlate data from pairs of detectors to search for such a Gaussian background \cite{Renzini:2023qtj}. 
Its likelihood rests on the hypothesis that the GWB is realized identically and independently across many segments~\cite{Allen:1999,Romano:2017}. 
The analysis reports an unbiased estimate of $S_h(f)$, combined optimally across frequency bins and segments under the Gaussian hypothesis.

Several dedicated searches have been developed that go beyond the Gaussian description.
For example: (i) Drasco and Flanagan~\cite{Drasco:2003} introduced a mixture-type likelihood, in which each segment is drawn either from noise alone or from signal-plus-noise, with probability $\xi$ that a given segment contains a signal.
(ii) Smith and Thrane~\cite{Smith:2018} extended~\cite{Drasco:2003} by treating each binary deterministically. 
They modeled the full event waveform $h(\theta)$ coherently and marginalized over its parameters to obtain the evidences for both signal-plus-noise and noise-only hypotheses.
This Bayesian analysis is optimal, but is potentially computationally intensive, requiring resolving and marginalizing over every potential event \cite{Kou:2025bhk}. 
(iii) Lawrence et al.~\cite{Lawrence:2023} introduced a stochastic-signal-based search, which models the per-segment signal not as a deterministic waveform but as a stochastic process, whose cross-correlated power is informed by the population through a hierarchical Bayesian prior.
The marginalization over the stochastic signal power using this prior is computationally less intensive than that of~\cite{Smith:2018} (see also \cite{Liu:2026}). Other proposals use fourth-order correlations of the data~\cite{Seto:2009ju} or a fourth-order Edgeworth expansion of the likelihood~\cite{Martellini:2014xia}, and Ref.~\cite{Buscicchio:2022raf} uses importance sampling of simulated signal realizations.

The non-Gaussian structure of the background has so far been studied as a property of the source population through the distribution of the strain and of its power~\cite{Meacher:2014aca,Ginat:2019aed,Lamb:2024gbh,Lamb:2025niq,Braglia:2022icu,Zhong:2026jzl,Bagui:2026hrf}.
However, it was unclear how to turn this description systematically into a search, since each search above assumes its own model for the signal in a segment.
In this Letter, we write the exact likelihood of the background for any rate and population in any network of detectors in terms of the characteristic function $z$ of the strain in a detector of a single merger.
This function follows directly from the population model and does not depend on the data.
Each approximation of $z$ defines a search with its own finite set of approximate sufficient statistics, and the cross-correlation search and the searches (i) to (iii) above follow as limiting cases.
Designing a search for a given population therefore reduces to approximating a single known function.
The goal then is to find those functions of the data that capture the most relevant information about a non-Gaussian GWB.

In supplemental material, we apply this formalism to an exactly solvable single-bin, toy-model example, and compare the detection floors and parameter recovery using the various likelihood function expansions. 
In a companion paper~\cite{Boybeyi:2026search}, we apply the same framework to a realistic population of binary black holes in a year of simulated data from the two LIGO detectors at O4 sensitivity. 
There, the framework gives a hierarchy of approximate sufficient statistics that starts from the cross-correlation and adds higher cross moments of the data at each order. 
This hierarchy of statistics allows us to construct a search that can detect the background from the subthreshold mergers alone, where the standard cross-correlation search does not.

\medskip
{\it Likelihood function}---\label{sec:likelihood}
The standard (Whittle) likelihood function for a GWB produced by the superposition of $K$ mergers in a data segment of duration $\tau$ is
\begin{equation}
\mathcal{L}\big(d\,\big|\,\{\lambda_i\}_{i=1}^{K}\big)
 = \prod_{I,\nu}\frac{1}{\pi\sigma_{I\nu}^{2}}\,
   \exp\!\left[-\frac{|d_{I\nu}-s_{I\nu}|^{2}}{\sigma_{I\nu}^{2}}\right],
\label{eq:like-data}
\end{equation}
where $d_{I\nu}$ denotes the complex Fourier amplitude of the data in detector $I$ and frequency bin $\nu$, and $s_{I\nu}$ denotes the total signal contribution
\begin{equation}
s_{I\nu} = \sum_{i=1}^{K} h_{I\nu}(\lambda_i)
\label{eq:ssum}
\end{equation}
from $K$ mergers.
Here, $h_{I\nu}(\lambda_i)$ is the strain produced by the $i$-th merger in the segment, with $\lambda_i$ denoting a set of parameters describing the $i$th merger.
The detector noise is assumed to be stationary-Gaussian with 1-sided noise power spectral density $S_{n_I}(f)$, which is related to the noise variance  $\sigma^2_{I\nu}$ in frequency bin $\nu$ via 
\begin{equation}
\sigma_{I\nu}^{2} \equiv \frac{\tau}{2}\,S_{n_I}(f_\nu)\,.
\label{eq:noisevar}
\end{equation}
In \eqref{eq:like-data}, the likelihood function is conditioned on the values of $\lambda_i$ for $i=1,2,\ldots, K$.

Since we are interested in the population properties of the sources contributing to the GWB, rather than the parameters of an individual merger, we marginalize the likelihood over the parameters of every merger and over the total number of mergers. 
Since the parameters for each merger are drawn independently from the population prior $\pi(\lambda_i|\Lambda)$, and the number of mergers $K$ is Poisson distributed with mean $\bar N$ (equal to
the merger rate times the segment duration), we obtain the marginalized segment likelihood:
\begin{widetext}
\begin{equation}
\mathcal{L}(d\,|\,\Lambda,\bar N)
 = \sum_{K=0}^{\infty}\frac{e^{-\bar N}\bar N^{K}}{K!}\int\!\prod_{i=1}^{K} d\lambda_i\,\pi(\lambda_i|\Lambda)\,
   \mathcal{L}\big(d\,\big|\,\{\lambda_i\}_{i=1}^{K}\big)\,.
\label{eq:margdef}
\end{equation}
The average over $K$ couples the mergers, since the exponent of \eqref{eq:like-data} is quadratic in the signal.

To perform the average one merger at a time, we make the exponent linear in the signal. 
We do this with the Fourier representation of the Gaussian, a Hubbard--Stratonovich transformation~\cite{Hubbard:1959ub}. 
More explicitly, for each $(I,\nu)$ we introduce a complex auxiliary variable $\eta_{I\nu}$ and use the Gaussian identity
\begin{equation}
\exp\!\left[-\frac{|d_{I\nu}-s_{I\nu}|^{2}}{\sigma_{I\nu}^{2}}\right]
 = \frac{\sigma_{I\nu}^{2}}{4\pi}\!\int\! d^{2}\eta_{I\nu}\,
   \exp\!\left[-\frac{\sigma_{I\nu}^{2}}{4}|\eta_{I\nu}|^{2}
   + i\,\mathrm{Re}\,\eta_{I\nu}^{*}\big(d_{I\nu}-s_{I\nu}\big)\right].
\label{eq:hubbard}
\end{equation}
The source now appears only linearly, so the exponential factorizes over the $K$ mergers, leading to 
\begin{equation}
\exp\!\bigg[-i\,\mathrm{Re}\!\sum_{I,\nu}\eta_{I\nu}^{*}s_{I\nu}\bigg]
 = \prod_{i=1}^{K}\exp\!\bigg[-i\,\mathrm{Re}\!\sum_{I,\nu}
   \eta_{I\nu}^{*}h_{I\nu}(\lambda_i)\bigg].
\label{eq:factorize}
\end{equation}
Since the identity~\eqref{eq:hubbard} is exact, the interference between overlapping merger waveforms has not been lost.

Each factor in square brackets on the rhs of~\eqref{eq:factorize} involves only a single merger event $i$, so the marginalization over its parameters $\lambda_i$ in~\eqref{eq:margdef} is also a single quantity per merger.
We denote this single-factor marginalization by
\begin{equation}
z(\eta;\Lambda)
 \equiv \big\langle
   \exp\!\big[-i\,\mathrm{Re}\textstyle\sum_{I,\nu}\eta_{I\nu}^{*}h_{I\nu}(\lambda)\big]
   \big\rangle_{\!\pi}\,.
   \label{eq:zdef}
\end{equation}
From its definition, we see that $z(\eta;\Lambda)$ has the form of a {\it characteristic function}%
\footnote{Recall that $\varphi_X(t)\equiv {\rm E}[e^{itX}]$ is the definition of the characteristic function of a single, real random variable $X$, which generalizes to multi-dimensional, complex-valued random variables as in \eqref{eq:zdef}.}
of a single merger's strain signal $h_{I\nu}(\lambda)$.
Since $z(\eta;\Lambda)$ is an average over the population prior of a known waveform, it does not depend on the data.
The Poisson sum over $K$ gives an exponential,
\begin{equation}
\sum_{K=0}^{\infty}\frac{e^{-\bar N}\bar N^{K}}{K!}\,
  \big[z(\eta;\Lambda)\big]^{K}
 = \exp\!\big[\bar N\,\big(z(\eta;\Lambda)-1\big)\big]\,.
\label{eq:poissonexp}
\end{equation}
Collecting~\eqref{eq:like-data} to~\eqref{eq:poissonexp}, the segment likelihood marginalized over the entire population takes the compact form
\begin{equation}
\mathcal{L}(d\,|\,\Lambda,\bar N)
 = \int\!\prod_{I,\nu} \frac{d^{2}\eta_{I\nu}}{4\pi^{2}}\;
   \exp\!\Bigg[
   -\sum_{I,\nu}\frac{\sigma_{I\nu}^{2}}{4}\,|\eta_{I\nu}|^{2}
   + i\,\mathrm{Re}\!\sum_{I,\nu}\eta_{I\nu}^{*}\,d_{I\nu}
   + \bar N\,\big(z(\eta;\Lambda)-1\big)
   \Bigg]\,.
   \label{eq:master}
\end{equation}
This is the main result of our paper and the master equation for the subsequent analyses.

\end{widetext}

Note that the population enters the integrand only via the argument of the exponential involving $z$, while the noise enters only via the Gaussian term in the auxiliary variables.
Since $z$ is the characteristic function of a single merger's signal, it equals one when $\eta$ goes to zero and its magnitude is bounded above by one, as expected for any characteristic function. 
For the continuous source populations that we consider, $z$ decreases as one moves away from the origin ($\eta=0$), as we shall see later. 

The noise factor of~\eqref{eq:master} confines the integral to $|\eta_{I\nu}| \lesssim 2/\sigma_{I\nu}$, while $z$ varies on the scale set by the strain of one merger.
So an expansion around the origin is accurate only when that strain is small compared to the noise. 
Thus, an accurate $z$ over the range that the noise allows generates accurate likelihoods, which makes $z$ a powerful target for approximation.

\medskip
{\it Limiting cases for the likelihood function}---\label{sec:limits}
We now show that the existing searches  for intermittent GWBs, which were briefly discussed in the introduction, follow from the master equation~\eqref{eq:master} as limiting expressions for the characteristic function. 
Note that we are free to: (i) expand $z$ around $\eta=0$, (ii) expand the Poisson exponent in the mean number of events per segment $\bar N$, or (iii) expand both $z$ and $\bar N$ simultaneously.

To expand $z$ around the origin $\eta=0$, we define $X\equiv\sum_{I,\nu}\eta_{I\nu}^{*}h_{I\nu}(\lambda)$ and make a Taylor series  expansion of the exponential in~\eqref{eq:zdef}.
This leads to
\begin{equation}
z(\eta;\Lambda) 
= 1 - i\,\big\langle \mathrm{Re}\,X \big\rangle_{\pi}- \frac{1}{2!}\big\langle (\mathrm{Re}\,X)^{2}\big\rangle_{\pi}
 + \cdots\,.
\label{eq:ztaylor}
\end{equation}
Since the waveform depends on the coalescence phase $\varphi_c$ via an overall complex factor, $h_{I\nu}(\lambda) = e^{i\varphi_c}\,\hat
h_{I\nu}(\hat\lambda)$ (where $\hat\lambda$ denotes the remaining parameters), the average of $h$ over $\varphi_c$ vanishes, implying $\langle{\rm Re}\, X\rangle_\pi=0$.
The second order term expands as $(\mathrm{Re}\,X)^{2} = (X^{2} + 2XX^{*} + X^{*2})/4$, where the averages $\langle hh\rangle_\pi$ and $\langle h^{*}h^{*}\rangle_\pi$ vanish for the same reason as above.
Only $\langle hh^{*}\rangle_\pi$ survives, yielding 
\begin{equation}
z(\eta;\Lambda)=
 1-\frac{1}{4}\sum_{I,\nu}\sum_{J,\nu'}\eta_{I\nu}^{*}\,
 \big\langle h_{I\nu}(\lambda)\,h_{J\nu'}^{*}(\lambda)\big\rangle_{\pi}\,\eta_{J\nu'}+\cdots\,.
\label{eq:zquad}
\end{equation}
The phase factor similarly removes every odd order, so the expansion of $z$ contains only even powers of $h$. 

(i) If we stop with the quadratic term in $h$ in~\eqref{eq:zquad} and substitute this truncated expansion for $z$ back into \eqref{eq:master}, we obtain an exponential of a quadratic expression in $\eta$.
The integral of this exponential is Gaussian, and can be evaluated by completing the square:
\begin{align}
&\mathcal{L}(d\,|\,\Lambda,\bar N)
= \frac{1}{\pi^{2N_f}\det C}\exp\!\big[-d^{\dagger}C^{-1}d\big]\,,\ {\rm with}
\label{eq:limcc}\\
&C_{I\nu,J\nu'}\equiv\delta_{IJ}\,\delta_{\nu\nu'}\,\sigma_{I\nu}^{2}
+ \bar N\,\big\langle h_{I\nu}(\lambda)\,h_{J\nu'}^{*}(\lambda)\big\rangle_{\pi}\,,
 \label{eq:ccov}
\end{align}
where $N_f$ is the number of (positive) frequency bins per detector and two detectors are assumed, the general normalization being $\pi^{N_{\rm det}N_f}$. 
This is the Gaussian likelihood of the standard cross-correlation search~\cite{Allen:1999,Matas:2021}.
The population enters it only through the rate (via $\bar N$) times the second moment of a single merger's waveform, which is the standard covariance matrix for a Gaussian GWB.

(ii) For small merger rate, the Poisson exponent expands to first order in
$\bar N$ as:
\begin{equation}
e^{\bar N (z-1)} \simeq \big(1-\bar N\big) + \bar N z.
\label{eq:rateexp}
\end{equation}
If we keep the full expression~\eqref{eq:zdef} for the characteristic function $z(\eta;\Lambda)$,  the $\eta$ integral of the signal term ``undoes" the identity of~\eqref{eq:hubbard} and recovers the likelihood~\eqref{eq:like-data} averaged over one merger:
\begin{widetext}
\begin{equation}
\mathcal{L}(d\,|\,\Lambda,\bar N)
 = \big(1-\bar N\big)\prod_{I,\nu}\frac{1}{\pi\sigma_{I\nu}^{2}}
   {\rm exp}\left[-\frac{|d_{I\nu}|^{2}}{\sigma_{I\nu}^2}\right]
   + \bar N\!\int\! d\lambda\,\pi(\lambda|\Lambda)
   \prod_{I,\nu}\frac{1}{\pi\sigma_{I\nu}^{2}}
   {\rm exp}\left[-\frac{|d_{I\nu}-h_{I\nu}(\lambda)|^{2}}{\sigma_{I\nu}^2}\right]\,.
\label{eq:limtbs}
\end{equation}
This is the marginalized mixture likelihood~\cite{Drasco:2003,Thrane:2013,Smith:2018} for a deterministic signal $h_{I\nu}(\lambda)$, with the duty cycle at leading order given by $\bar N$.

(iii) Finally, the stochastic-based search for intermittent backgrounds (SSI)~\cite{Lawrence:2023} can be recovered by keeping~\eqref{eq:rateexp}, but replacing the quadratic expansion~\eqref{eq:zquad} of $z(\eta,\lambda)$ by its exponential form:
\begin{equation}
z(\eta;\Lambda) \simeq \exp\!\Big[-\frac{1}{4}\sum_{I,\nu}\sum_{J,\nu'}\eta_{I\nu}^{*}\,
 \big\langle h_{I\nu}(\lambda)\,h_{J\nu'}^{*}(\lambda) \big\rangle_{\pi}\,\eta_{J\nu'}\Big]\,.
\label{eq:zgauss}
\end{equation}
The above expression for $z(\eta,\Lambda)$ agrees with~\eqref{eq:zquad} up to a correction at fourth-order in $\eta$.
Both terms of~\eqref{eq:rateexp} then give Gaussian
integrals, which can be evaluated to yield:
\begin{equation}
\mathcal{L}(d\,|\,\Lambda,\bar N)
 = \big(1-\bar N\big)\prod_{I,\nu}\frac{1}{\pi\sigma_{I\nu}^{2}}
   {\rm exp}\left[-\frac{|d_{I\nu}|^{2}}{\sigma_{I\nu}^2}\right]
   + \bar N\,\frac{1}{\pi^{2N_f}\det C_{1}}\exp\!\big[-d^{\dagger}C_{1}^{-1}d\big]\,,
\label{eq:limssi}
\end{equation}
where $C_{1}$ is the covariance of~\eqref{eq:ccov} for $\bar N=1$.
This is the SSI likelihood given in~\cite{Lawrence:2023} at fixed segment power; the marginalization of that power over the population prior in~\cite{Lawrence:2023} amounting to an average of $z$ in \eqref{eq:zgauss} and of the Gaussian term in \eqref{eq:limssi} over that prior. 

Each of these three searches follows from a different truncation of the same characteristic function, which raises the question of what other approximations
are available.

\medskip
{\it Families of approximate likelihoods and sufficient statistics}---
We show now that expanding $z(\eta;\Lambda)$ (or $\bar{N}(z(\eta;\Lambda)-1)$) in a basis of functions of $\eta$ defines a family of approximate likelihoods, in which the data enter only through the combinations (statistics) that the basis generates. 
Since $z(\eta;\Lambda)$ does not depend on data and detector noise properties, the basis is chosen for a given population and applies to any analysis. 
Choosing the most efficient statistics is then the same problem as approximating the characteristic function with the fewest terms.

We begin this approximation procedure by noting that the likelihood function~\eqref{eq:master} can be written as 
\begin{equation}
{\mathcal{L}(d\,|\,\Lambda,\bar N)}={\mathcal{L}_0(d)}\,\big\langle \exp\!\big[\bar N\big(z(\eta;\Lambda)-1\big)\big]
   \big\rangle_{\eta}\,,
   \label{eq:noiseavg}
\end{equation}
where $\mathcal{L}_0(d)$ is the pure noise likelihood,
\begin{equation}
\mathcal{L}_0(d)
= \prod_{I,\nu}\frac{1}{\pi\sigma^2_{I\nu}}
\exp\left[-\frac{|d_{I\nu}|^2}{\sigma^2_{I\nu}}\right]
 \equiv \int\!\prod_{I,\nu} \frac{d^{2}\eta_{I\nu}}{4\pi^{2}}\;
   \exp\!\Bigg[
   -\sum_{I,\nu}\frac{\sigma_{I\nu}^{2}}{4}\,|\eta_{I\nu}|^{2}
   + i\,\mathrm{Re}\!\sum_{I,\nu}\eta_{I\nu}^{*}\,d_{I\nu}
   \Bigg]\,,
   \label{eq:noiseonly}
\end{equation}
and $\langle\exp\left[\bar N(z(\eta;\Lambda)-1)\right]\rangle_\eta$ is the population factor averaged over $\eta$, where $\langle\ \rangle_\eta$ is defined by
\begin{equation}
\langle f(\eta)\rangle_\eta
\equiv \left[\mathcal{L}_0(d)\right]^{-1}\,\int\!\prod_{I,\nu} \frac{d^{2}\eta_{I\nu}}{4\pi^{2}}\;
   \exp\!\Bigg[
   -\sum_{I,\nu}\frac{\sigma_{I\nu}^{2}}{4}\,|\eta_{I\nu}|^{2}
   + i\,\mathrm{Re}\!\sum_{I,\nu}\eta_{I\nu}^{*}\,d_{I\nu}
   \Bigg]f(\eta_{I\nu})\,.
   \label{eq:etaavg}
\end{equation}

We then assume that we can expand the argument of the exponential in~\eqref{eq:noiseavg} in a set of basis of functions
\begin{equation}
\bar N\big(z(\eta;\Lambda)-1\big)\simeq \sum_{n} w_n(\Lambda,\bar N)\,\psi_n(\eta)\,,
\label{eq:expexp}
\end{equation}
where $w_n(\Lambda,\bar N)$ denote the weights of this expansion, and the approximate equality means that this expansion need not be complete.
Inserting this expansion into~\eqref{eq:noiseavg}, we obtain
\begin{equation}
{\mathcal{L}(d\,|\,\Lambda,\bar N)}={\mathcal{L}_0(d)}\,\Big\langle \exp\!\Big[\sum_n w_n(\Lambda,\bar N)\,
   \psi_n(\eta)\Big]\Big\rangle_{\eta}\,.
\label{eq:insertphi}
\end{equation}
Note that the data enter through the $\eta$-averaged expectations of the basis functions, which we denote as
\begin{equation}
S_n(d) \equiv \big\langle \psi_n(\eta) \big\rangle_{\eta}\,.
\label{eq:tdef}
\end{equation}
The logarithm of~\eqref{eq:insertphi} then expands in the cumulants of the basis functions,
\begin{equation}
\ln\frac{{\mathcal{L}(d|\Lambda,\bar N)}}{\mathcal{L}_0(d)}
 =\sum_n w_n(\Lambda,\bar N)\,S_n(d)
 +\frac12\sum_{m,n} w_m(\Lambda,\bar N)\,
   w_n(\Lambda,\bar N)\big[\big\langle\psi_m\psi_n\big\rangle_{\eta}
   -S_m(d)\,S_n(d)\big]+\cdots\,.
\label{eq:cumexp}
\end{equation}
For many segments, the likelihood of the analysis is the product over the segments. Summing \eqref{eq:cumexp} over them:
\begin{equation}
\ln\prod_i\frac{\mathcal{L}(d_i\,|\,\Lambda,\bar N)}{\mathcal{L}_0(d_i)}
=\Nseg\Big[\sum_n w_n(\Lambda,\bar N)\,\bar S_n
+\frac12\sum_{m,n}w_m(\Lambda,\bar N)\,w_n(\Lambda,\bar N)\big[\overline{\langle\psi_m\psi_n\rangle_\eta}-\overline{S_mS_n}\big]+\cdots\Big]\,,
\label{eq:tbar}
\end{equation}
with $\bar S_n(d)\equiv\Nseg^{-1}\sum_{i=1}^{\Nseg}S_n(d_i)$, where the overbar denotes an average over segments.

Note that $\bar S_n$ are single-merger statistics, while the averages of the quadratic combinations $\overline{\langle\psi_m\psi_n\rangle_\eta}-\overline{S_mS_n}$ are two-merger statistics.
One sees this by first re-expressing the $\sum_n w_n(\Lambda,\bar N)\bar S_n$ factors in \eqref{eq:tbar} in terms of $z(\eta;\Lambda)$ using \eqref{eq:expexp}, and then writing the $\eta$-averages of $z(\eta;\Lambda)$ in terms of the $\lambda$-averages of $h_{I\mu}(\lambda)$ using \eqref{eq:zdef}.
For example, the two-merger statistics contain expressions like
\begin{equation}
\big\langle z^{2}(\eta;\Lambda)\big\rangle_{\eta}
 = \int\! d\lambda_1\, d\lambda_2\,
\pi(\lambda_1|\Lambda)\,\pi(\lambda_2|\Lambda)\Big\langle\exp\left[-i\,\mathrm{Re}\textstyle\sum_{I,\nu}
   \eta_{I\nu}^{*}[h_{I\nu}(\lambda_1)+h_{I\nu}(\lambda_2)]\right]
   \Big\rangle_{\eta}\,.
   \label{eq:zsq}
\end{equation}
Each order of \eqref{eq:tbar} thus builds statistics for one more overlapping merger, the confusion noise of the background entering the search through them rather than through a model of its spectrum.

In practice, we truncate \eqref{eq:tbar} after the first-order term in $w_n$, which keeps the single-merger statistics and drops the two-merger ones, whose data-independent part returns below as the normalization $A(w)$.
For each segment, we then have
\begin{equation}
\label{eq:1storder}
\mathcal{L}(d\,|\,\Lambda,\bar N)
\approx \mathcal{L}_0(d)\exp\left[ \sum_n w_n(\Lambda,\bar N)\, S_n(d)\right]\,.
\end{equation}
Because of the truncation, the right-hand side of \eqref{eq:1storder} is not normalized.
One approach to normalizing the truncated likelihood is to numerically integrate the right-hand side of \eqref{eq:1storder} over the data to obtain the segment normalization factors $\mathcal{N}$, and then divide the right-hand side by these factors.
Alternatively, we can explicitly introduce an additional term $A(w)$ in the likelihood to enforce the normalization:
\begin{equation}
\label{eq:1storder_with_A}
\mathcal{L}(d\,|\,\Lambda,\bar N)
\approx \mathcal{L}_0(d)\exp\left[ \sum_n w_n(\Lambda,\bar N)\, S_n(d)-A(w)\right]\,,
\end{equation}
where $\int\mathcal{D}d\,\mathcal{L}(d\,|\,\Lambda,\bar N)=1$ implies $A(w)=\ln\int\mathcal{D}d\,\mathcal{L}_0(d)\,\exp\big[\sum_n w_n S_n(d)\big]$.
Expanding the exponential in this expression and using $\int\mathcal{D}d\,\mathcal{L}_0(d)\,\langle f(\eta)\rangle_\eta=f(0)$ (which is valid for any $f(\eta)$), we find that the first-order term $\sum_n w_n\psi_n(0)$ vanishes by \eqref{eq:expexp} at $\eta=0$, while the second-order term gives
\begin{equation}
A(w)\simeq \frac{1}{2}\sum_{m,n} w_m w_n\,
\mathrm{Cov}_0\big[S_m(d),S_n(d)\big]\,.
\label{e:norm}
\end{equation}
Using this expression for $A(w)$, the truncated likelihood is normalized up to $O(w^3)$.
We can estimate ${\rm Cov}_0$ directly from the data, under the assumption that signal is weak compared to the noise, or calculate it analytically for Gaussian noise parametrized, e.g., by $\sigma_{I\nu}$. 
Both approaches give approximate expressions for $A(w)$.

We can go further in the limit of many segments, where the averages $\bar S_n$ are Gaussian by the central-limit theorem and their likelihood depends on the population only through the mean and covariance of $S_n(d)$.
This form also avoids the segment-level normalization $\int\mathcal{D}d\,\mathcal{L}(d\,|\,\Lambda,\bar N)=1$, which is not defined for bases whose statistics $S_n(d)$ grow faster than quadratically in the data, since $\exp[\sum_n w_nS_n(d)]$ is then not integrable.
With large number of segments we can write
\begin{equation}
\label{eq:emplike}
\ln\mathcal{L}(\bar S\,|\,\Lambda,\bar N)
\simeq-\frac{\Nseg}{2}\sum_{m,n}(\bar S_m-\mu_m)[\Sigma^{-1}]_{mn}(\bar S_n-\mu_n)
-\frac{1}{2}\ln\det(\Sigma/\Nseg)\,,
\end{equation}
where
\begin{equation}
\mu_n(\Lambda,\bar N)\equiv\int\mathcal{D}d\,\mathcal{L}(d\,|\,\Lambda,\bar N)\,S_n(d)\,,
\qquad
\Sigma_{mn}(\Lambda,\bar N)\equiv\int\mathcal{D}d\,\mathcal{L}(d\,|\,\Lambda,\bar N)\,S_m(d)\,S_n(d)-\mu_m\mu_n\,.
\end{equation}
Note that $\mu_n$ and $\Sigma_{mn}$ depend on both $\bar N$ and the population parameters $\Lambda$.
For $\bar N=0$, these reduce to $\mu_n=\psi_n(0)$ and $\Sigma_{mn}=\mathrm{Cov}_0[S_m(d),S_n(d)]$, which depend only on the noise and can be estimated from the data as above, while their dependence on $\Lambda$ and $\bar N$ follows from the population model, for example through its analytic moments.
\end{widetext}

Alternatively, to fully leverage the non-Gaussianity of the individual events for improved probability of detection, we could instead presume that the merger rate is small and use the mixture form of~\eqref{eq:rateexp}.
We then apply the same expansion to the characteristic function of one merger's signal,
\begin{equation}
z(\eta;\Lambda) \simeq \sum_n w_n(\Lambda)\,\psi_n(\eta)\,,
\label{eq:zexp}
\end{equation}
where the coefficients $w_n$ no longer depend on $\bar N$.
The $\eta$ integral of~\eqref{eq:master} then gives a mixture in every segment,
\begin{equation}
\mathcal{L}(d\,|\,\Lambda,\bar N)
 = \big(1-\bar N\big)\,\mathcal{L}_0(d)
 + \bar N\,\mathcal{L}_0(d) \sum_n w_n(\Lambda)\,S_n(d),
\label{eq:mixfamily}
\end{equation}
where $S_n(d) \equiv\langle \psi_n(\eta)\rangle_{\eta}$ are the per-segment statistics. This family generalizes the mixture searches of~\cite{Drasco:2003,Thrane:2013,Smith:2018,Lawrence:2023}, while the exponential family~\eqref{eq:1storder_with_A} generalizes the cross-correlation search. 

Note, however, that this form \eqref{eq:mixfamily} sacrifices the data compression property of the exponential family. Namely, the data terms in the joint likelihood over many segments cannot be combined into the averages given in~\eqref{eq:tbar}, and the analysis keeps the statistics of every segment.

Therefore, we see from these two expansions that the likelihood of the analysis can be written, at every truncation, as a function of a finite set of approximate sufficient statistics, which are different for every choice of the basis. 
Hence, there is a one-to-one correspondence between the representations of the characteristic function $z(\eta;\Lambda)$ of~\eqref{eq:zdef} and the families of approximate sufficient statistics of the search.

\medskip
{\it Conclusion}---
\label{sec:conclusion}
In this letter, we asked the question: which functions of the data carry the most relevant information about a non-Gaussian GWB. 
To answer that question, we reformulated the detection problem in terms of the characteristic function of a single merger's strain signal in multiple detectors, and we showed that the current known searches are different limits of it, obtained by approximating the characteristic function about the origin.
Each approximation generates a family of approximate sufficient statistics, so our reformulation is a concrete way to explicitly construct them. 

In the Supplemental Material, we apply our formalism to an exactly solvable single-bin toy model, and show that one statistic beyond the cross-correlation leverages the intermittent nature of the background making the merger rate and per-event power separately measurable.
In a companion paper Ref.~\cite{Boybeyi:2026search}, we build such a search for two separated detectors that recovers the merger rate and the mass distribution of a realistic population from a year of simulated data at O4 sensitivity.

\medskip
\noindent
\textit{Acknowledgments:}  We are grateful to Arianna Renzini and the LVK Stochastic Group for valuable comments. TB and VM were in part supported by the NSF grant PHY-2409173. 
JDR acknowledges support from NSF grant PHY-2207270 and start-up funds at the University of Texas Rio Grande Valley.

\bibliography{refs}


\onecolumngrid
\clearpage
\appendix

\section{\large Supplemental material:\\
Constructing likelihoods for the non-Gaussian gravitational-wave background}

\section{\large I. Definition of single-bin toy model}
\label{sec:singlebin}
We demonstrate our ``constructing-likelihoods" framework on the following toy model example, which corresponds to a single-frequency bin observed by two co-located and co-aligned detectors with real-valued data.
We choose this example because the exact likelihood and the characteristic function are both analytically known, which a fully realistic case does not allow.
This means that the approximations below can then be tested against the exact answer. 

Each detector observes 
the common signal plus its own independent realization of Gaussian noise,
\begin{equation}
d_I=h+n_I,\qquad n_I\sim\mathcal{N}(0,\sigma^2),\qquad I=1,2,
\label{eq:toymodel}
\end{equation}
and the signal is a sum of a Poisson number of unresolved events,
\begin{equation}
h=\sum_{k=1}^{K}h_k,\qquad K\sim\mathrm{Poisson}(\bar N),\qquad h_k\sim\mathcal{N}(0,A_0^2),
\label{eq:toysignal}
\end{equation}
with each event contributing a real amplitude shared by both detectors. The two model parameters are the mean number of mergers per-segment $\bar N$ and the per-event power $A_0^2$.

The auxiliary variables for this example consist of a single real-valued $\eta_I$ per detector---i.e., $\eta\equiv (\eta_1,\eta_2)$, since the data~\eqref{eq:toymodel} are real-valued strain samples and \eqref{eq:hubbard} needs only one real variable per strain sample. 
The characteristic function $z$ defined in~\eqref{eq:zdef} is the marginalization over the strain $h$ for a single event.
Since the marginalization for this example is a Gaussian integral, we can evaluate it analytically:
\begin{equation}
z(\eta;A_0)=\int\frac{dh}{\sqrt{2\pi}\,A_0}\,e^{-h^2/(2A_0^2)}\,
e^{-i(\eta_1+\eta_2)\,h}=\exp\!\Big[-\tfrac12 A_0^2\,(\eta_1+\eta_2)^2\Big].
\label{eq:toyz}
\end{equation}
Note that for this case, the difference $\eta_1-\eta_2$ does not contain information about the population of events, since the two detectors share the same signal and $z$ depends on $\eta$ only through the sum $\eta_1+\eta_2$.

The Poisson sum of Eq.~\eqref{eq:poissonexp} then makes the likelihood an exact mixture of Gaussians, one for each event count,
\begin{equation}
\mathcal{L}(d\,|\,\bar N,A_0)=e^{-\bar N}\sum_{K=0}^{\infty}\frac{\bar N^{K}}{K!}\,
\frac{1}{2\pi\,\sigma\sqrt{\sigma^2+2KA_0^2}}\,
\exp\!\Big[-\frac{(d_1+d_2)^2}{4(\sigma^2+2KA_0^2)}-\frac{(d_1-d_2)^2}{4\sigma^2}\Big]\,,
\label{eq:toyexact}
\end{equation}
where $d\equiv(d_1,d_2)$.
This is the exact likelihood for this single-bin case, and each term is the contribution of a fixed event count $K$. 
The variance of the sum $d_1+d_2$ grows linearly with the number of events, while the difference $d_1-d_2$ is independent of $K$ and thus remains pure noise. 
The zero-event term is the pure-noise likelihood $\mathcal{L}_0$, and the one-event term is the single-event correction. 
The rate, via $\bar N$, weights the sum, so the first few terms dominate when $\bar N$ is small, where the background departs most from a Gaussian.

\section{\large II. Approximate likelihoods}
\label{sec:approx}

We now construct three families of approximate likelihoods for this model. Each family extends one of the searches described in the main text:
(i) The event count truncation extends the marginalized mixture search of~\cite{Drasco:2003,Thrane:2013,Smith:2018}; 
(ii) the Gaussian mixture approximation extends the stochastic-signal based inference of~\cite{Lawrence:2023}, and (iii) the Taylor-series approximation extends the standard cross-correlation search of LVK analyses~\cite{Allen:1999}.

\subsection{II.A. Event count truncation approximation}
\label{s:event_count}

The exact mixture likelihood~\eqref{eq:toyexact} already defines a family of approximate likelihoods, before any expansion of the characteristic function. 
The approximation here is on the event count alone and the characteristic function enters exactly via \eqref{eq:toyz}. 
So, here we simply truncate the event-count sum at $K\le n$:
\begin{equation}
\mathcal{L}_{K\le n}(d\,|\,\bar N,A_0)\equiv
\frac{\displaystyle\sum_{K=0}^{n}\frac{\bar N^{K}}{K!}\,\mathcal{L}_K(d|\bar N,A_0)}
{\displaystyle\sum_{K=0}^{n}\frac{\bar N^{K}}{K!}}\,,
\label{eq:Ktrunc}
\end{equation}
where 
\begin{equation}
\mathcal{L}_K(d\,|\,\bar N, A_0)
\label{eq:toyLK}
\equiv 
   \frac{1}{2\pi\sigma\sqrt{\sigma^{2}+2KA_0^{2}}}\,
   \exp\!\Big[-\frac{(d_1+d_2)^{2}}{4(\sigma^{2}+2KA_0^{2})}
   -\frac{(d_1-d_2)^{2}}{4\sigma^{2}}\Big]
   \end{equation}
is the exact likelihood for a segment containing $K$ mergers.
We have included the summation in the denominator of \eqref{eq:Ktrunc} to renormalize the retained Poisson weights in the numerator, which is needed to have a properly normalized probability distribution for the data.
The truncation family of likelihood functions converges to the exact likelihood as $n$ increases.

What the truncation approximation drops is the chance that more than $n$ mergers land in one segment.
This is negligible at the low merger rates of an astrophysical background.
Note that $\mathcal{L}_{K\le 1}$ is the ``popcorn'' limit of a single loud merger, corresponding to both \eqref{eq:limtbs} and \eqref{eq:limssi}.
These likelihoods are identical for this single-bin toy model, since $h_k\sim \mathcal{N}(0,A_0)$ is a random variable.

\subsection{II.B.  Gaussian mixture approximation}
\label{s:gaussian_mixture}

For our second approximate expansion, we use a family of Gaussian basis functions,
\begin{equation}
\psi_n(\eta) = \exp\!\Big[-\tfrac12 C_n\,(\eta_1+\eta_2)^{2}\Big],
\label{eq:gaussbasis}
\end{equation}
where the scales $C_n$ are fixed numbers that we determine later.
We then expand the population factor of~\eqref{eq:noiseavg} in terms of this basis,
\begin{equation}
\exp\!\big[\bar N\big(z(\eta)-1\big)\big]
 \simeq \sum_{n} w_n(\bar N, A_0)\,\psi_n(\eta)\,,
\qquad w_n \ge 0\,,
\label{eq:mmansatz}
\end{equation}
which leads to the likelihood
\begin{equation}
\mathcal{L}(d\,|\,\bar N, A_0)
 = \mathcal{L}_0(d) \sum_{n} w_n(\bar N, A_0)\, S_n(d),
\label{eq:mmlike}
\end{equation}
where
\begin{equation}
S_n(d) \equiv \big\langle \psi_n(\eta) \big\rangle_{\eta}= \sqrt{\frac{\sigma^{2}}{\sigma^{2}+2C_n}}
   \exp\!\left[\frac{(d_1+d_2)^{2}}{4}
   \Big(\frac{1}{\sigma^{2}}
   -\frac{1}{\sigma^{2}+2C_n}\Big)\right]\,.
\label{eq:mmstat}
\end{equation}
These statistics are just the likelihood ratios of the Gaussian components. 
Each term of~\eqref{eq:mmlike} integrates to one over the data, so we need $\sum_n w_n = 1$ for a proper normalization. 

The member with one constant scale and one scale at the single-merger covariance $A_0^2$ coincides with both the SSI and marginalized mixture likelihoods,~\eqref{eq:limssi} and~\eqref{eq:limtbs}, since the single-event strain amplitude of this model is Gaussian and the two searches agree for this case. 
The family extends these likelihoods to multiple Gaussians in the signal part with fitted weights.

\begin{figure*}[h!]
\centering
\includegraphics[width=0.95\textwidth]{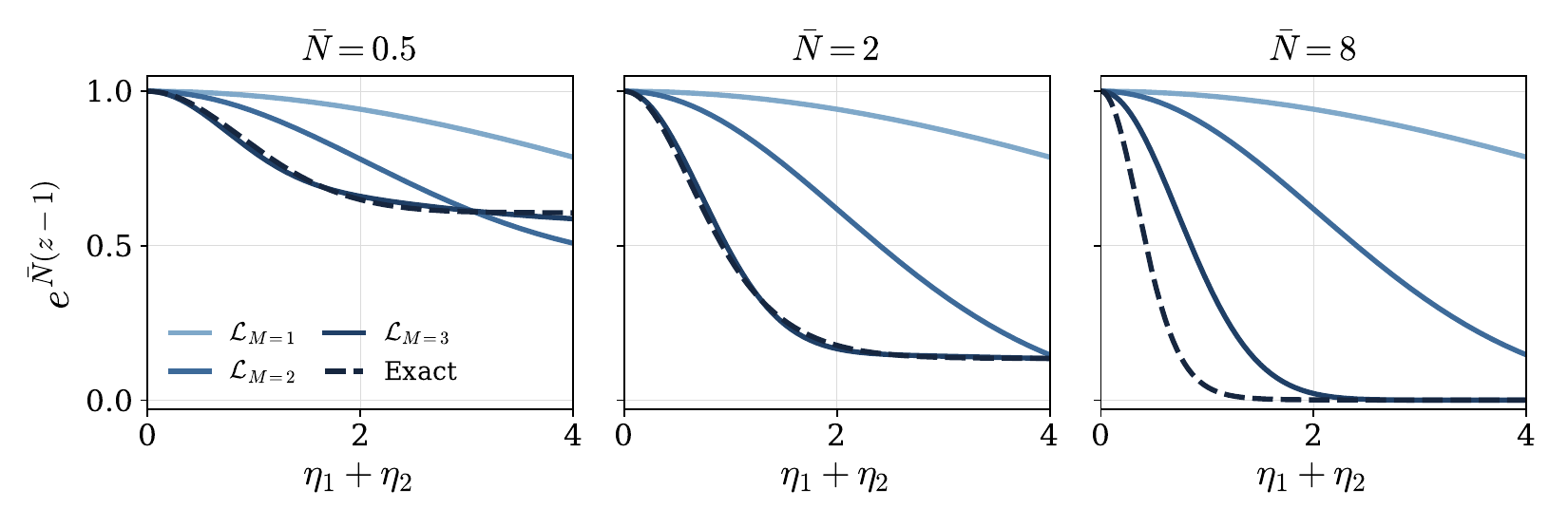}
\caption{Plots of $\exp[\bar N(z-1)]$ as a function of $\eta_1+\eta_2$, along with the approximations used in the mixture members $\mathcal{L}_{M=1}$, $\mathcal{L}_{M=2}$, $\mathcal{L}_{M=3}$, for different values of $\bar N$ at unit per-event power $A_0^2=1$. 
As the number of Gaussian components grows, the mixture converges to the true
characteristic function. 
Each member of this likelihood family is therefore an approximation to the exact likelihood.}
\label{fig:momentz}
\end{figure*}

The scales can in general be tied to the detectors through $\sigma^2$, or to the population, as in the exact mixture of~\eqref{eq:toyexact}, where they are $n A_0^2$. 
Here we take $C_n = c_n \sigma^2$ with fixed numbers $c_n$, since $\sigma^2$ is estimated from the data and the basis is then independent of the population parameters. 
We choose $c_n = \{0, 0.03, 0.24, 1.92\}$, which is not optimized and could be tuned for a given population. 
The component of scale $C_n$ corresponds to segments whose signal power is of that order.
Thus, the constant component corresponds to segments with no events, while the largest scale corresponds to segments with several events. 
The weights $w_n(\bar N, A_0)$ follow from a least squares fit of $\exp[\bar N(z-1)]$ with $w_n \ge 0$ and $\sum_n w_n = 1$, and the member $\mathcal{L}_{M}$ keeps at most $M$ components of nonzero scale, the constant component with $c_0=0$ being always present, different from zero. 
Figure~\ref{fig:momentz} shows the fit as the number of components grows.

\subsection{II.C. Taylor series approximation}
\label{s:taylor}

Our final family of approximate likelihoods is obtained by approximating the characteristic function in~\eqref{eq:toyz} via a Taylor series expansion about the origin.
This family extends the cross-correlation search, which keeps just the quadratic statistic. 
Expanding the exact expression for $z(\eta; A_0)$ in~\eqref{eq:toyz} in powers of $A_0^2$ gives
\begin{equation}
z(\eta;A_0)= \exp\left[-\frac{1}{2}\, A_0^2(\eta_1+\eta_2)^2\right]= \sum_{n\ge 0} \frac{1}{n!}
   \Big(-\tfrac12 A_0^2\Big)^{n} (\eta_1+\eta_2)^{2n}\,.
\label{eq:toyexp}
\end{equation}
The basis functions and the weights of the expansion~\eqref{eq:expexp} of $\bar N(z-1)$ are thus
\begin{equation}
\psi_n(\eta) = (\eta_1+\eta_2)^{2n}\,,
\qquad
w_n(\bar N,A_0) = \frac{\bar N}{n!}\Big(-\tfrac12 A_0^2\Big)^{n}\,,\qquad n\ge1\,.
\label{eq:toygphi}
\end{equation}
Averaging the basis functions over the Gaussian weight of~\eqref{eq:master}, which gives $\eta_1+\eta_2$ the mean $i\,s/\sigma^2$ and the variance $2/\sigma^2$ with $s\equiv d_1+d_2$, gives the statistics
\begin{equation}
\begin{aligned}
S_1(d) &= \frac{2\sigma^2 - s^2}{\sigma^4}\,,\\
S_2(d) &= \frac{s^4 - 12\sigma^2 s^2 + 12\sigma^4}{\sigma^8}\,,\\
S_3(d) &= \frac{-s^6 + 30\sigma^2 s^4 - 180\sigma^4 s^2 + 120\sigma^6}{\sigma^{12}}\,.
\label{eq:toystats}
\end{aligned}
\end{equation}
$S_1$ is the cross-correlation (and auto-correlation) statistic, while $S_2$ and $S_3$ carry the departure of the data from a Gaussian background.

Conditional on the event count, $s$ is Gaussian with variance $2\sigma^2+4KA_0^2$ (see~\eqref{eq:toyLK}) so the moments follow in closed form from a Poisson average over $K$.
With $r\equiv A_0^2/\sigma^2$, the means are
\begin{equation}
\mu_1=-\frac{4\bar N r}{\sigma^2}\,,\qquad
\mu_2=\frac{48\,(\bar N+\bar N^2)\,r^2}{\sigma^4}\,,\qquad
\mu_3=-\frac{960\,(\bar N+3\bar N^2+\bar N^3)\,r^3}{\sigma^6}\,,
\label{eq:toymu}
\end{equation}
and, for example, $\Sigma_{11}=(8/\sigma^4)[1+4\bar Nr+6\bar Nr^2+4\bar N^2r^2]$.
For $\bar N=0$ they reduce to $\mu_n=0$ and $\Sigma_{mn}=\mathrm{Cov}_0[S_m(d),S_n(d)]=\delta_{mn}(2/\sigma^2)^{2n}(2n)!$.
The fourth cumulant of $s$ follows from the means as $\kappa_4(s)/\sigma^8=\mu_2-3\mu_1^2$.

The members $\mathcal{L}_2$, $\mathcal{L}_4$, $\mathcal{L}_6$ are~\eqref{eq:emplike} with the first 1, 2, and 3 statistics, given in~\eqref{eq:famthree}.

\medskip
To summarize, we collect the three families for easy comparison:
\begin{align}
\mathcal{L}_{K\le n}(d\,|\,\bar N,A_0)
 &= \Bigg[\sum_{K=0}^{n}\frac{\bar N^{K}}{K!}\Bigg]^{-1}
\sum_{K=0}^{n}\frac{\bar N^{K}}{K!}\,
   \frac{1}{2\pi\sigma\sqrt{\sigma^{2}+2KA_0^{2}}}\,
   \exp\!\left[-\frac{(d_1+d_2)^{2}}{4(\sigma^{2}+2KA_0^{2})}
   -\frac{(d_1-d_2)^{2}}{4\sigma^{2}}\right]\,,
\label{eq:famone}\\
\mathcal{L}_{M}(d\,|\,\bar N,A_0)
 &= \sum_{n} w_n(\bar N,A_0)\,
   \frac{1}{2\pi\sigma\sqrt{\sigma^{2}+2C_n}}\,
   \exp\!\left[-\frac{(d_1+d_2)^{2}}{4(\sigma^{2}+2C_n)}
   -\frac{(d_1-d_2)^{2}}{4\sigma^{2}}\right]\,,
\label{eq:famtwo}\\
\mathcal{L}_{2m}(\bar S(\{d_i\})\,|\,\bar N,A_0)
 &= \frac{1}{\sqrt{(2\pi/\Nseg)^{m}\det\Sigma}}\,
   \exp\!\left[-\frac{\Nseg}{2}\sum_{k,l=1}^{m}\big(\bar S_k-\mu_k\big)\big[\Sigma^{-1}\big]_{kl}\big(\bar S_l-\mu_l\big)\right]\,,
\label{eq:famthree}
\end{align}
with $w_n$ in~\eqref{eq:famtwo} the fitted weights of~\eqref{eq:mmansatz}, $\bar S_n$ in~\eqref{eq:famthree} the segment averages of the statistics~\eqref{eq:toystats}, and $\mu_n$, $\Sigma_{mn}$ their moments from Sec.II.C.

\section{\large III. Assessing the performance of the likelihoods}
\label{s:assessment}

We assess the performance of the three families of approximate likelihoods based on their respective detection floors (defined below) and their parameter recovery. 
We draw the event count $K$ of a segment from the Poisson distribution of~\eqref{eq:toysignal}, we draw the $K$ strain amplitudes $h_k$ from a Gaussian of width $A_0$, and we add an independent noise draw from a Gaussian of width $\sigma$ in each detector. 
The difference channel $d_1 - d_2$ has no signal contribution in this model, so in every case we estimate the noise variance $\sigma^2$ from it, and use that estimate in evaluating likelihoods of~\eqref{eq:famone} to~\eqref{eq:famthree}. 
Note that the real analysis follows the same logic. 
The LVK search estimates the noise power spectra of an analysis segment from the averaged periodograms of the neighboring segments, so the estimate shares no data with the statistic it weights, and the residual signal contribution is negligible in the noise-dominated regime~\cite{Romano:2017,Matas:2021}.

We first choose a grid of rates $\bar N$ and per-event powers $A_0^{2}$. 
For each grid point we simulate an observation of $4\times10^{4}$ segments of data having those parameters, and we repeat it for $5\times10^{3}$ realizations. 
We also simulate $5\times10^{3}$ realizations of data consisting of noise alone. 
For each $\bar N$ we evaluate the log likelihood ratio of each approximate likelihood, maximized over a grid of the two population parameters, both for the data realizations containing events and for those containing only noise.
The threshold is set on the noise realizations at the false alarm probability ${\rm FAR} = 0.1$, and we increase $A_0^{2}$ until the detection rate of the realizations with events reaches $0.9$.
Finally we mark that power, which gives the detection floor $\bar N A_0^{2}$ of that likelihood at that rate. 
We then construct the floor of every likelihood over the whole rate
grid, and the recovery contours at three benchmark populations:  the maximum likelihood point and the $\Delta\ln\mathcal{L} = -3$ contour of every realization over $(\log_{10}\bar N, \log_{10}A_0^{2})$ with the truth marked, as seen in Figs.~\ref{fig:mixcallout}, \ref{fig:mmcallout}, and~\ref{fig:clcallout}. 

\section{\large IV. Results}
\label{sec:results}

\begin{figure*}[p]
\centering
\includegraphics[width=0.9\textwidth]{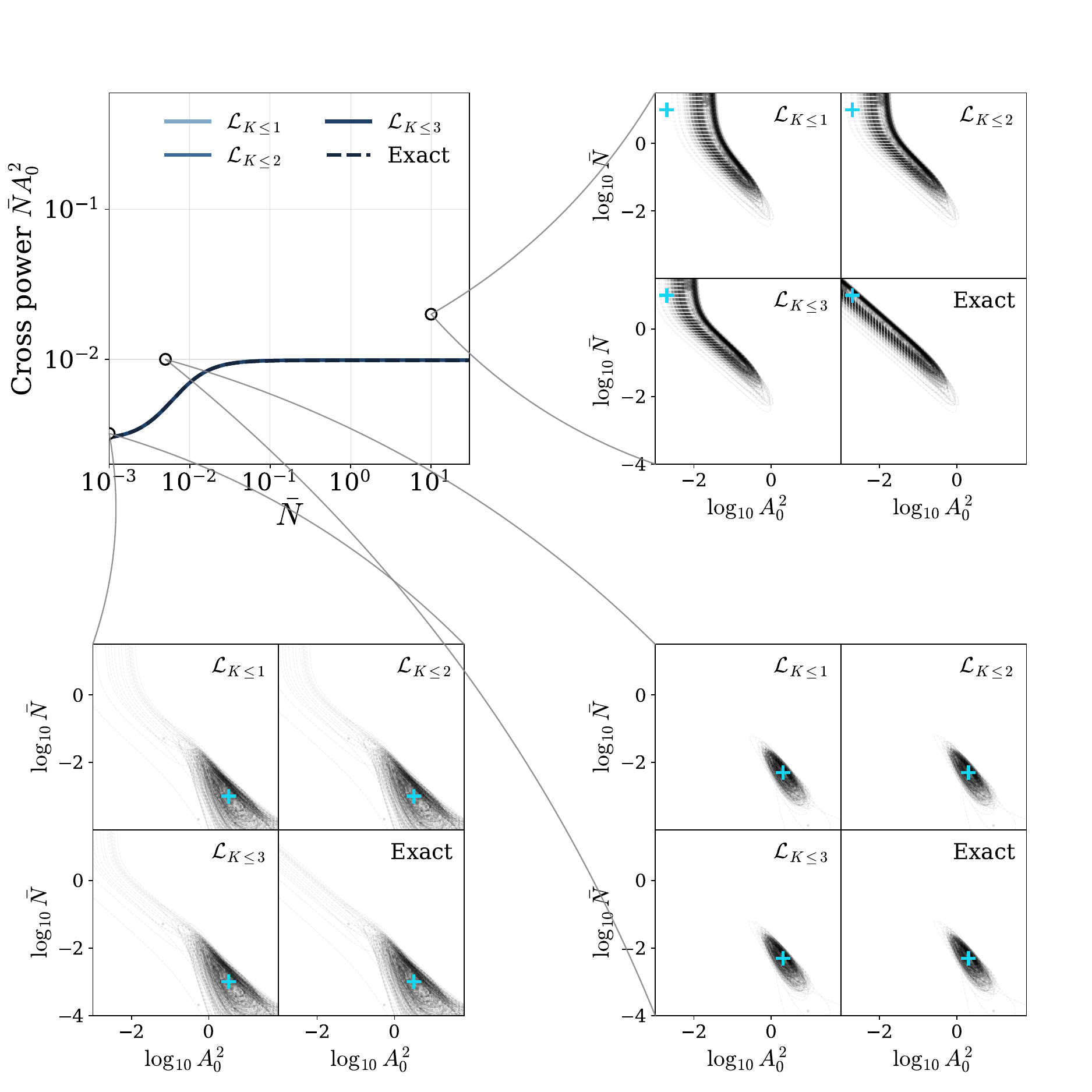}
\caption{Detection and recovery for the event-count family of~\eqref{eq:Ktrunc}, for a Gaussian intermittent signal in two co-located and co-aligned detectors. 
The top-left panel is the smallest detectable cross power $\bar N A_0^2$ versus the rate $\bar N$ for $\mathcal{L}_{K\le1}$, $\mathcal{L}_{K\le2}$, $\mathcal{L}_{K\le3}$ and the
exact mixture; this is the detection floor of each likelihood. 
The callouts give, at the three marked rates, the maximum-likelihood recovery of $(\log_{10}\bar N,\log_{10}A_0^2)$. 
Each thin curve is the $\Delta\ln\mathcal{L}=-3$ contour of one realization, and the cyan cross marks the  true parameter values. 
Where the background is rare and loud, the one-event truncation already reaches the exact detection floor and centers on the true parameter values, while at higher rates the low truncations stay on the exact floor and bias the estimate.
The family closes in on the exact mixture as more event counts are kept.}
\label{fig:mixcallout}
\end{figure*}

\begin{figure*}[p]
\centering
\includegraphics[width=0.9\textwidth]{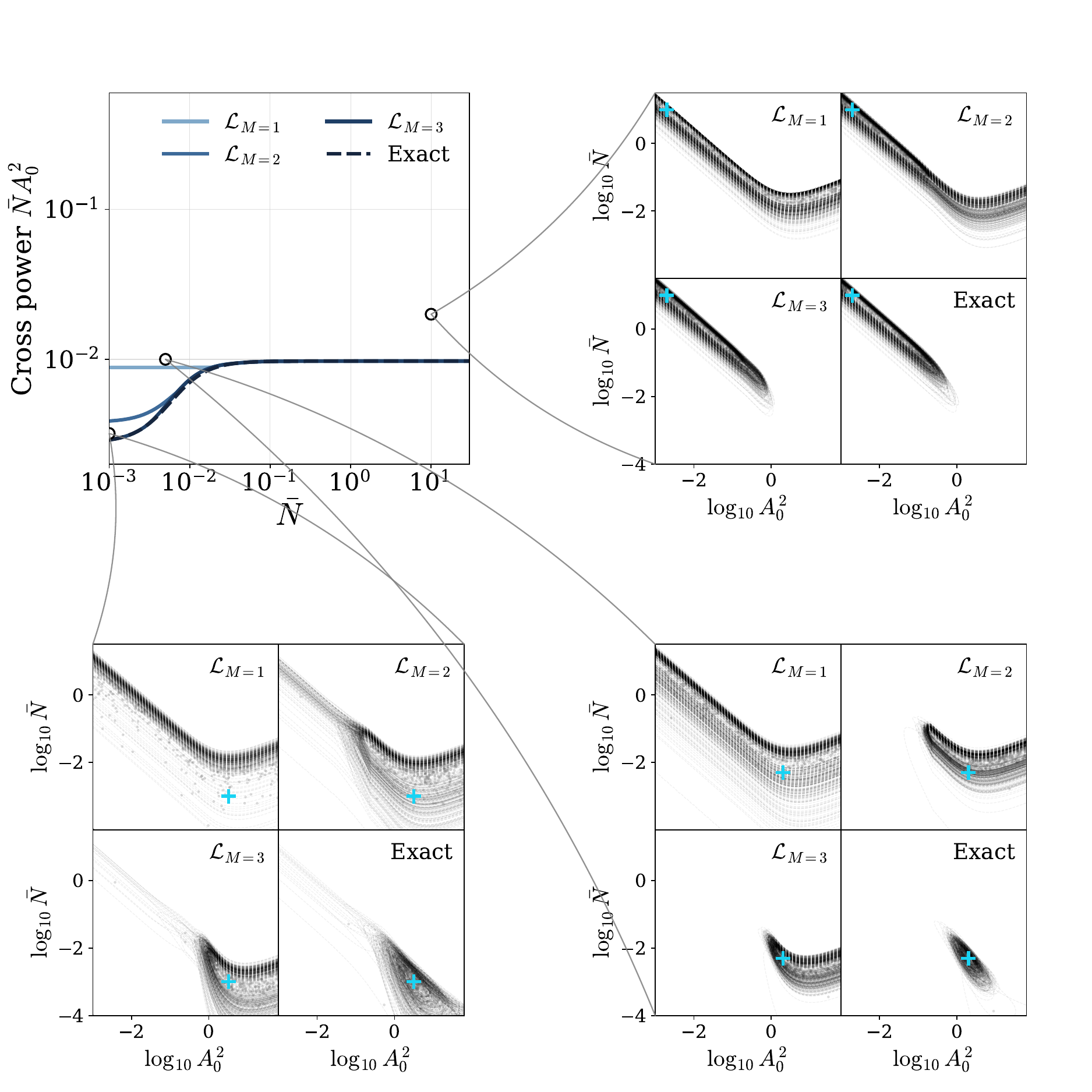}
\caption{Detection and recovery for the Gaussian-mixture family of~\eqref{eq:mmansatz}, in the same setup as Fig.~\ref{fig:mixcallout}. 
The top-left panel is the smallest detectable cross power $\bar N A_0^2$ versus the rate $\bar N$ for
$\mathcal{L}_{M=1}$, $\mathcal{L}_{M=2}$, $\mathcal{L}_{M=3}$ and the exact mixture. The callouts give, at the three marked rates, the maximum-likelihood recovery of $(\log_{10}\bar N,\log_{10}A_0^2)$. 
Each thin curve is the $\Delta\ln\mathcal{L}=-3$ contour of one realization and the cyan cross marks the true parameter values. 
The posterior from the single-component likelihood $\mathcal{L}_{M=1}$ is informative only on the degenerate combination $\bar N A_0^2$, and its detection floor is above that of the exact likelihood. Each additional component narrows the posterior and lowers the detection floor toward that of the exact mixture.}
\label{fig:mmcallout}
\end{figure*}

\begin{figure*}[p]
\centering
\includegraphics[width=0.9\textwidth]{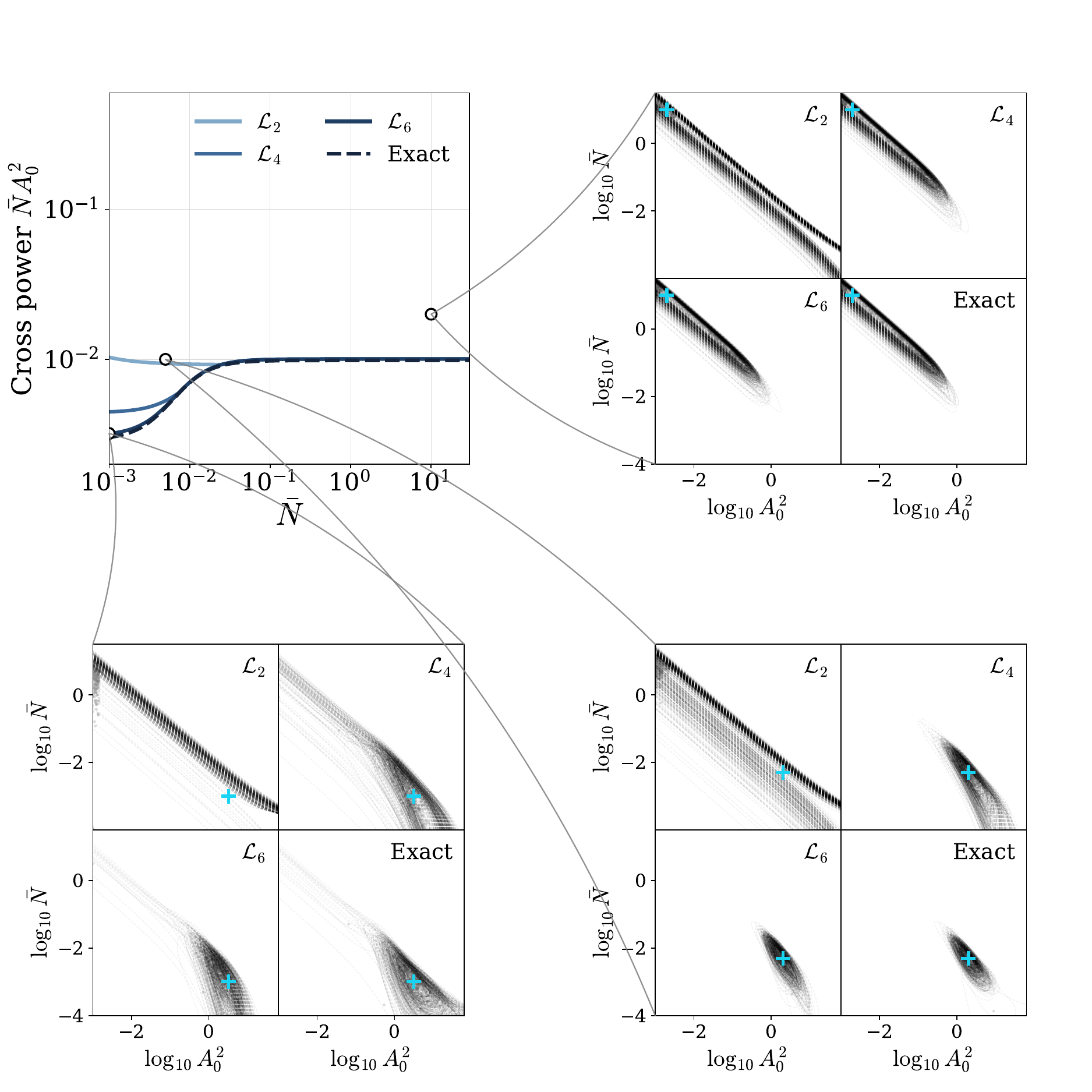}
\caption{Detection and recovery for the Taylor family of~\eqref{eq:famthree}, in the same setup as Fig.~\ref{fig:mixcallout}.
The top-left panel is the smallest detectable cross power $\bar N A_0^2$ versus the rate $\bar N$ for $\mathcal{L}_2$, $\mathcal{L}_4$, $\mathcal{L}_6$ and the exact mixture. 
The callouts give, at three benchmark rates, the maximum-likelihood recovery of $(\log_{10}\bar N,\log_{10}A_0^2)$.
Each thin curve is the $\Delta\ln\mathcal{L}=-3$ contour of one realization and the cyan cross marks the true parameter values.
Each added data power tightens the recovery and lowers the floor toward the exact mixture.}
\label{fig:clcallout}
\end{figure*}

The top left panels of Figs.~\ref{fig:mixcallout}, \ref{fig:mmcallout}, and~\ref{fig:clcallout} show the detection floors of the three families, and the dashed curve shows the floor of the exact likelihood~\eqref{eq:toyexact}. 
The exact floor flattens at $\bar N A_0^{2} \simeq 10^{-2}$ for $\bar N \gtrsim 10^{-2}$, where the background is effectively Gaussian, and it drops by about a factor of three towards $\bar N = 10^{-3}$. 
A background of rare and loud events is therefore easier to detect than one with the same total power in many quiet events, and the non-Gaussian search improves on the Gaussian search for $\bar N \lesssim 10^{-2}$.

We start with the $\mathcal{L}_{K\le n}$ family in Fig.~\ref{fig:mixcallout}. 
The floors of $\mathcal{L}_{K\le1}$, $\mathcal{L}_{K\le2}$, and $\mathcal{L}_{K\le3}$ lie on top of each other and on the exact floor at every rate. 
At low rate, a segment contains at most one event, and the chance of a second event is smaller by another factor of $\bar N$.
So the higher event counts correct the likelihood at order $\bar N^{2}$, and the floor does not resolve them. 
At high rate, a segment contains many events and the likelihoods with small $K$ are not the correct model. 
Their floors still flatten at the same level because: (i) the floor the signal in every segment is weak, (ii) the log likelihood ratio of each of them then reduces at leading order to the same quadratic statistic, and (iii) the detection power follows from that statistic alone.

The other two families reach the exact floor with a growing number of terms, as shown in Figs.~\ref{fig:mmcallout} and~\ref{fig:clcallout}. 
The floors of $\mathcal{L}_{M=1}$ and $\mathcal{L}_{2}$ are flat over most of the range, because a single Gaussian and the quadratic statistic involve the total power alone, up to the dependence of $\Sigma_{11}$ on $\bar N A_0^{4}$, and the floor of $\mathcal{L}_2$ rises slightly below $\bar{N}\simeq 10^{-2}$. 
The quadratic statistic averages $(d_1+d_2)^2$ over the segments, and at fixed cross power the variance of that average gains a term proportional to $\bar N A_0^{4}$.
So at low rate, its scatter is set by the few loud segments rather than by the noise. $\mathcal{L}_{M=2}$ and $\mathcal{L}_{4}$ recover most of the non-Gaussian advantage, and $\mathcal{L}_{M=3}$ and $\mathcal{L}_{6}$ converge further.

The $\Delta\ln\mathcal{L} = -3$ contours illustrate the same point in the model-parameter domain. 
We choose three benchmark populations, marked on the floor panels, and every thin contour in the callouts is the $\Delta\ln\mathcal{L} = -3$ contour of one realization. 
At the $\bar N = 10$ benchmark, only the total power is measurable, and the exact likelihood gives a straight degenerate contour along constant $\bar N A_0^{2}$ through the true parameter values. 
At high rate the contours of the event-count likelihoods deviate from the injected values, because a truncation at $n$ events per segment distributes the same total power over fewer and louder events, and the deviation persists for $\mathcal{L}_{K\le 3}$.
The detection floors of these likelihoods are unchanged, so the truncation biases the parameter recovery at high rate and leaves the detection unaffected.

At $\bar N = 10^{-3}$ and $\bar N = 5\times10^{-3}$, the signal part of the $2m$-th cumulant of $d_1+d_2$ is proportional to $\bar N A_0^{2m}$.
So the line on which it is constant grows steeper with $m$, from slope $-1$ for the second cumulant, to $-2$ and $-3$ for the fourth and the sixth. 
$\mathcal{L}_{2}$ measures the second cumulant alone, up to the same dependence of $\Sigma_{11}$, and its contour stays open along the least steep of these lines, while $\mathcal{L}_{4}$ and $\mathcal{L}_{6}$ narrow it across the steeper ones. 
$\mathcal{L}_{M=1}$, $\mathcal{L}_{M=2}$, $\mathcal{L}_{M=3}$ show a similar hierarchy where the larger scales $C_n$ are more sensitive to higher power of the data.

At the $\bar N = 10$ benchmark the event-count and the Gaussian-mixture families both depart from the degenerate line of constant $\bar N A_0^{2}$, for different reasons. 
The contours of $\mathcal{L}_{K\le1}$, $\mathcal{L}_{K\le2}$ and $\mathcal{L}_{K\le3}$ are displaced from the truth, since the truncation constrains the per event power rather than leaving it degenerate. 
The contour of $\mathcal{L}_{M=1}$ contains the true parameter values, but is not single valued in $A_0^{2}$, since the member depends on the population through a single weight, $\mathcal{L}_{M=1} = \mathcal{L}_0(d)[(1-w_1) + w_1 S_1(d)]$, and $w_1(\bar N, A_0)$ is non-monotonic in $A_0^{2}$. 
The same behavior is weaker in $\mathcal{L}_{M=2}$ and absent in $\mathcal{L}_{M=3}$. 
At $\bar N = 10^{-3}$, the true parameter values lie below the floors of $\mathcal{L}_{M=1}$ and $\mathcal{L}_{2}$, so neither approximate likelihood detects it and their contours remain unbounded.

The efficient statistics are the representations that capture the exact likelihood with the fewest terms. 
The Gaussian-mixture family reaches the exact floor at three components, since the population factor is a flat part and a decay that a few Gaussians represent over the range of $\eta$ that the noise allows, see Fig.~\ref{fig:momentz}, although three components do not reproduce the full Poisson mixture of \eqref{eq:toyexact}. 
The Taylor family needs three statistics for the detection floor and for parameter recovery in the transition region, and two already recover most of the non-Gaussian advantage.

\end{document}